\documentclass[reprint,superscriptaddress,prx] {revtex4-1}
\usepackage{graphicx}% Include figure files
\usepackage{hyperref}
\hypersetup{hidelinks}
\usepackage{color}
\usepackage{autobreak}
\usepackage{dcolumn}
\usepackage{verbatim}

\usepackage{xcolor}
\usepackage{xpatch}
\makeatletter
\def\changeBibColor#1{%
  \in@{#1}{}%  list of colored bib items
  \ifin@\color{red}\else\normalcolor\fi
}
 
\xpatchcmd\@bibitem
  {\item}
  {\changeBibColor{#1}\item}
  {}{\fail}
 
\xpatchcmd\@lbibitem
  {\item}
  {\changeBibColor{#2}\item}
  {}{\fail}

\usepackage{textcomp}
\begin{document}

%\preprint{APS/123-QED}

\title{Assessing Phonon Coherence Using Spectroscopy}% Force line breaks with \\

\author{Zhongwei Zhang}
\affiliation{Center for Phononics and Thermal Energy Science, School of Physics Science and Engineering and China-EU Joint Lab for Nanophononics, Tongji University, Shanghai 200092, People's Republic of China}
\affiliation{Institute of Industrial Science, The University of Tokyo, Tokyo 153-8505, Japan}

\author{Yangyu Guo}
\affiliation{Institut Lumière Matière, Université Claude Bernard Lyon 1-CNRS, Université de Lyon, Villeurbanne 69622, France}

\author{Marc Bescond}
\affiliation{IM2NP, UMR CNRS 7334, Aix-Marseille Université, Faculté des Sciences de Saint Jérôme, Case 142, 13397 Marseille Cedex 20, France}

\author{Masahiro Nomura}
\affiliation{Institute of Industrial Science, The University of Tokyo, Tokyo 153-8505, Japan}

\author{Sebastian Volz}
\email{volz@iis.u-tokyo.ac.jp}
\affiliation{Center for Phononics and Thermal Energy Science, School of Physics Science and Engineering and China-EU Joint Lab for Nanophononics, Tongji University, Shanghai 200092, People's Republic of China}
\affiliation{Laboratory for Integrated Micro and Mechatronic Systems, CNRS-IIS UMI 2820, The University of Tokyo, Tokyo 153-8505, Japan}

\author{Jie Chen}
\email{jie@tongji.edu.cn}
\affiliation{Center for Phononics and Thermal Energy Science, School of Physics Science and Engineering and China-EU Joint Lab for Nanophononics, Tongji University, Shanghai 200092, People's Republic of China}

\date{\today}

\begin{abstract}

As a fundamental physical quantity of thermal phonons, temporal coherence participates in a broad range of thermal and phononic processes, while a clear methodology for the measurement of phonon coherence is still lacking. In this Lettter, we derive a theoretical model for the experimental exploration of phonon coherence based on spectroscopy, which is then validated by comparison with Brillouin light scattering data and direct molecular dynamic simulations of confined modes in nanostructures. The proposed model highlights that confined modes exhibit a pronounced wavelike behavior characterized by a higher ratio of  coherence time to lifetime. The dependence of phonon coherence on system size is also demonstrated from spectroscopy data. The proposed theory allows for reassessing data of conventional spectroscopy to yield coherence times, which are essential for the understanding and the estimation of phonon characteristics and heat transport in solids in general.

\end{abstract}

\maketitle

Coherence is a fundamental characteristic of the wavelike behavior of elementary particles and quasi-particles \cite{ZH606,ZH608,RN1438,zhang2021coherent}. Due to the  wave nature of lattice vibrations, phonon coherence has been established as the dominant source of various unique thermal transport phenomena, such as coherent thermal transport (including minimum thermal conductivity in nanophononic crystals \cite{Simkin2000,ZH606,ZH608,RN289,ZH1230,ZH975} and phonon Anderson localization in disordered nanophononic crystals \cite{Mendoza2016,Luckyanova2018,Juntunen2019,Hu2021,Guo2021}), band folding \cite{Tamura1999,Bies2000,Simkin2000} and phonon confinement \cite{Gupta2009,RN194,Lee2017}. Furthermore, recent studies \cite{RN1616,RN1617,Zhang2021,Jain2020,RN1617,zhang2021strong,Jain2020,RN1617,zhang2021strong} have uncovered the significant impact of phonon coherence on phonon-phonon scattering \cite{RN1616,RN1617,Zhang2021}, phonon modal correlations \cite{Jain2020,RN1617,zhang2021strong} and interfacial phonon propagation \cite{Hu2021}, leading to clear discrepancies with the predictions obtained from the particle picture and also to promote coherence as a critical phonon attribute. 

Phonon coherence is usually quantified by the physical quantity of coherence time \cite{RN1815}. However, most of the experimental demonstrations of coherence for thermal phonons are qualitatively inferred from the variation of measured thermal conductivity \cite{ZH606,ZH608,Hu2020,Luckyanova2018}. For instance, the existence of coherent phonons is indirectly evidenced by the experimental observation of the non-monotonic variation of superlattice thermal conductivity with decreasing superlattice period \cite{ZH608}. On the other hand, phononic quantities, including phonon number, eigenfrequency and scattering rate, can be measured from the well-established time-domain thermoreflectance approach \cite{Jiang2017,Zhu2017,MacCabe2020} or various spectroscopies \cite{Burkel2000,Serrano2007,Minnich2011,Glensk2019,Zacharias2021,ruan2022}, which are boosting the exploration of phonon physics and thermal transport in different matters. Clearly, the experimental characterization of the coherence of thermal phonons is lagging behind, due to the lack of an adequate theory, which impedes the scientific understanding and practical application of phonon physics and thermal transport.

In this Letter, we develop a theoretical model for experimentally investigating phonon coherence using spectroscopy measurements. The predictions of the model disclose that in the case of confined nanostructures, such as nanowires (NWs) and nanomembranes (NMs), the coherence time for different modes can be efficiently extracted. Because of the substantial phonon confinement, modes have a prominent wavelike characteristic and exhibit a significant dimensionality-dependent coherence. Furthermore, the phonon coherence times and lifetimes detected from spectroscopy agree qualitatively well with the outcomes from the direct molecular dynamic (MD) simulations. The proposed theory reassesses mainstream spectroscopies to quantify mode coherence having a general impact on the understanding and the estimation of phonon and thermal properties.
 
Phonon lifetime is conventionally measured from the Lorentzian fitting of experimental spectra, such as inelastic neutron scattering, inelastic x-ray scattering and Brillouin-Mandelstam light scattering \cite{Glensk2019,RN1414,RN2004,RN2005,RN194}. The underlying physics of a Lorentzian fit is based on the hypothesis of an exponential decay of phonon dynamics in time, assuming phonons as particles governed by diverse scattering processes \cite{ZH72,RN1533}. In the frequency domain, the Lorentzian function is expressed as \cite{Ladd1986,RN489,ZH72}

\begin{eqnarray}
{\Phi _{\mathbf{k}s}}\left ( \omega  \right )=\frac{I_{0}}{4\left ( \omega-\omega_{\mathbf{k}s}  \right )^{2}{\tau _{\mathbf{k}s}^{p}}^{2}+1},
\label{eq1}
\end{eqnarray}

\noindent where ${\Phi _{\mathbf{k}s}}\left ( \omega  \right )$ is the spectral mode energy and $I_{0}$ is the peak intensity. The physical quantities of eigenfrequency ($\omega_{\mathbf{k}s}$) and lifetime ($\tau _{\mathbf{k}s}^{p}$) are simultaneously evaluated by matching the experimental measurements as implemented in Refs. \cite{Glensk2019,RN194}. The Eq. (\ref{eq1}) can also be expressed in terms of the modal linewidth ($\gamma _{\mathbf{k}s}$) according to the expression of $\tau _{\mathbf{k}s}^{p}=\frac{1}{2\gamma _{\mathbf{k}s}}$. Using the normal mode decomposition method, the spectral energy ${\Phi _{\mathbf{k}s}}\left ( \omega  \right )$ can also be calculated from MD simulations, making of Eq. (\ref{eq1}) a widely used model for studying the phonon lifetimes and thermal transport in theoretical and numerical simulation works \cite{RN489,ZH72,zhang2018reducing,Zhang2020a,RevModPhys.94.025002}.

Recent studies \cite{Zhang2021,zhang2021heat,zhang2021strong} demonstrated the significance of coherence on the phonon dynamics and thermal transport in various solids. By taking into account the coherence effects, a pioneering work \cite{Zhang2021} established that the time-dependent phonon number should be corrected as follows

\begin{eqnarray}
N_{\mathbf{k}s}\left ( t \right )=N_{\mathbf{k}s}\left ( 0 \right ) e^{-\gamma _{\mathbf{k}s}'t}  e^{-4ln2\cdot \Omega_{\mathbf{k}s}^{2}t^{2}},
\label{eq2}
\end{eqnarray}

\noindent where $N_{\mathbf{k}s}\left ( 0 \right )$ denotes the phonon number of mode $\mathbf{k}s$ at the initial time of decay. $\gamma'$ represents the corrected linewidth for this mode and $\Omega_{\mathbf{k}s}$ denotes the inverse of the temporal extension of the wavepacket. The corrected phonon lifetime from Eq. (\ref{eq2}) writes $\tau _{\mathbf{k}s}^{p'}=\frac{1}{2\gamma _{\mathbf{k}s}'}$ and the temporal coherence time $\tau _{\mathbf{k}s}^{c}=\frac{1}{\Omega_{\mathbf{k}s}}$.

On the other hand, the phonon number can also be defined from the normal mode coordinate $\mathbf{q}_{\mathbf{k}s}$ as

\begin{eqnarray}
N_{\mathbf{k}s}\left ( t \right )\hbar\omega_{\mathbf{k}s}=\frac{1}{2}\dot{\mathbf{q}}_{\mathbf{k}s}(t)\dot{\mathbf{q}}^{\ast }_{\mathbf{k}s} (t)+\frac{1}{2}\omega_{\mathbf{k}s}\mathbf{q}_{\mathbf{k}s}(t)\mathbf{q}^{\ast }_{\mathbf{k}s} (t),
\label{eq3}
\end{eqnarray}

\noindent where $\ast$ indicates the complex conjugate. Based on Eqs. (\ref{eq2}) and (\ref{eq3}), we can further infer the expression of the normal mode coordinate as \cite{RN1815}

\begin{eqnarray}
\mathbf{q}_{\mathbf{k}s}(t) \approx q_{0} e^{- i \omega_{\mathbf{k}s} t} e^{- \frac{\gamma_{\mathbf{k}s}'}{2}t} e^{-2ln2\cdot \Omega_{\mathbf{k}s}^{2}t^{2}},
\label{eq4}
\end{eqnarray}

\noindent where $q_{0} $ denotes the amplitude and $e^{- i \omega_{\mathbf{k}s} t} $ corresponds to the natural oscillation of the lattice wave. In a further step, the mode energy can be obtained from the Fourier transform of Eq. (\ref{eq4}) as

\begin{eqnarray}
\Phi _{\mathbf{k}s}\left ( \omega  \right )=2\times \frac{1}{2} \omega_{\mathbf{k}s}\left | \int_{0}^{\infty } \mathbf{q}_{\mathbf{k}s}(t) e^{i\omega t}dt\right |^{2}.
\label{eq5}
\end{eqnarray}

\noindent The coefficient 2 accounts for the summation of the kinetic and the potential terms. Finally, the spectroscopy model that includes both lifetime and coherence time is obtained as

\begin{eqnarray}
\Phi _{\mathbf{k}s}\left ( \omega  \right )=Icos\left [ \frac{1}{16ln2} \frac{ {\tau _{\mathbf{k}s}^{c2}} }{\tau _{\mathbf{k}s}^{p'}} \left ( \omega-\omega_{\mathbf{k}s}  \right ) \right ]e^{\frac{- \left ( \omega-\omega_{\mathbf{k}s}  \right )^{2}{\tau _{\mathbf{k}s}^{c2}}}{8ln2}},
\label{eq6}
\end{eqnarray}

\iffalse
\begin{flalign}
\Phi _{\mathbf{k}s}\left ( \omega  \right )=Icos\left [ \frac{\pi }{8ln2} \frac{ {\tau _{\mathbf{k}s}^{c2}} }{\tau _{\mathbf{k}s}^{p}} \left ( \omega-\omega_{\mathbf{k}s}  \right ) \right ]e^{\frac{-\pi^{2}\left ( \omega-\omega_{\mathbf{k}s}  \right )^{2}{\tau _{\mathbf{k}s}^{c2}}}{2ln2}} .
\label{eqn2}
\end{flalign}
\fi

\noindent where $I$ denotes the peak intensity. In contrast to the Lorentzian model of Eq. (\ref{eq1}), our theory can simultaneously provide the information of eigenfrequency ($\omega_{\mathbf{k}s} $), coherence corrected phonon lifetime ($\tau _{\mathbf{k}s}^{p'}$) and coherence time ($\tau _{\mathbf{k}s}^{c}$). 

As previously discussed, the spectral energy can be measured by experimental spectroscopy as implemented by many different techniques, and can also be calculated from the normal mode decomposition with inputs from MD simulations. However, Eq. (\ref{eq6}) reveals that $\tau _{\mathbf{k}s}^{p'}$ and $\tau _{\mathbf{k}s}^{c}$ are coupled in the spectral domain, making the fitting for these two quantities non-unique. 

A previous study \cite{Zhang2021} showed that the lifetimes fitted by Eq. (\ref{eq1}) are close to the coherence corrected ones, i.e., $\tau _{\mathbf{k}s}^{p}\approx \tau _{\mathbf{k}s}^{p'}$. Following this approximation, the lifetimes and coherence times can be estimated from the spectral energy via a two-step fitting: 1) the lifetime and eigenfrequency can be obtained by fitting the spectral energy with Eq. (\ref{eq1}); 2) the second fitting is performed to extract the coherence time from Eq. (\ref{eq6}) based on 1). The fitting code has been implemented in our open-source WPPT package \cite{code}, which is originally designed to study the coherence of phonons and the thermal conductivity of solids.

\begin{figure}[t]
\includegraphics[width=1.0\linewidth]{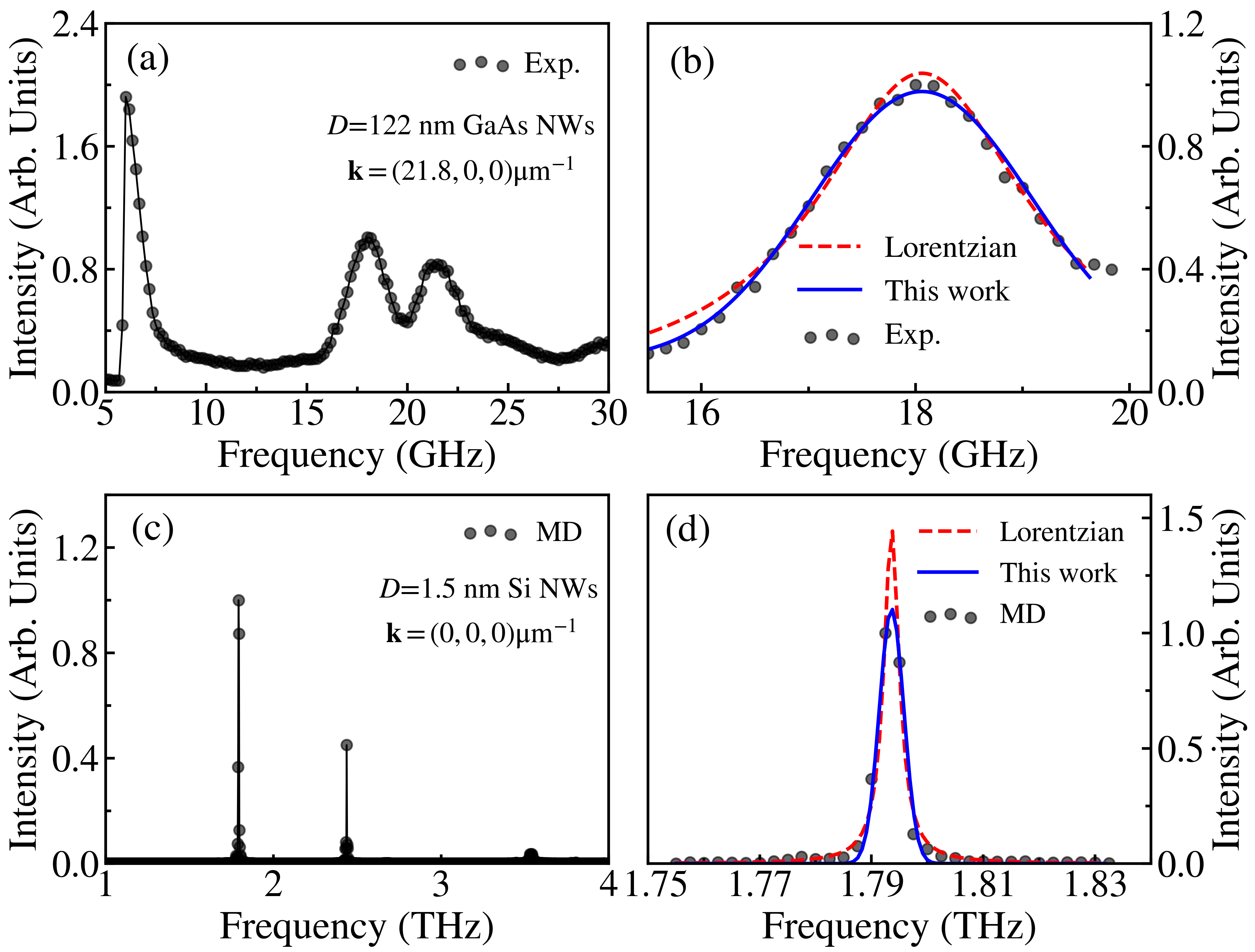}
\caption{(a) The experimental spectrum of GaAs nanowires with a diameter $D=122$ nm at the specific wavevector $\mathbf{k}=$ (21.8,0,0) \textmu m$^{-1}$. (b) The fitting of experimental spectroscopy data with the classical Lorentzian model of Eq. (\ref{eq1}) and the proposed one in Eq. (\ref{eq6}). The experimental data were previously published by Kargar $et$ $al.$ \cite{RN194}. (c) The MD spectral analysis of Si nanowires with a diameter $D=1.5$ nm at the specific wavevector $\mathbf{k}=$ (0,0,0) \textmu m$^{-1}$. (d) The fitting of the MD calculated spectral energy with the classical Lorentzian model of Eq. (\ref{eq1}) and the one of Eq. (\ref{eq6}).
}
\label{fig1}
\end{figure}

To support the proposed model, we further investigate the coherence of phonons in the confined nanostructures. Previous studies \cite{Gupta2009,Crowe2011,RN194,Lee2017,RN1829} demonstrated that phonon confinement appears along the non-periodic direction as reducing the structure dimensionality to one or two. In one-dimensional NWs, for example, phonons are confined in the diameter direction as standing waves and exhibit a diameter-dependent confinement \cite{RN194,Adu2006,Zhu2017}. For instance, Kargar $et$ $al.$ \cite{RN194} have measured the phonon subbands in GaAs NWs using the Brillouin-Mandelstam light scattering spectroscopy. Figure \,\ref{fig1}(a) reports the experimental spectrum from Ref. \cite{RN194} at a specific wavevector for a GaAs NW with a diameter $D=122$ nm. Peaks corresponding to different phonon modes clearly appear. The fitting of a specific mode is shown in Fig.\,\ref{fig1}(b). The proposed model displays a better agreement with the spectroscopy data than the Lorentzian fit does. To provide further comparison, we also calculate the spectral energy of Si NWs with smaller diameters, from MD simulations. The details about MD simulations and the normal mode decomposition are provided in Supplementary Material \cite{SM}. Figure\,\ref{fig1}(c) reports the spectral energy at wavevector $\mathbf{k}=$ (0,0,0) for a Si NW with a diameter $D=1.5$ nm. The fitting is carried out for a specific mode in Fig.\,\ref{fig1}(d) and the proposed model was found to provide a high accuracy as found in the experimental spectroscopy data of Fig.\,\ref{fig1}(b), when the coherence effects are considered in the full phonon dynamics. The same model is also applied to study the phonon coherence in two-dimensional NMs, as reported in Sec. S3 of the Supplementary Material \cite{SM}. Figure S2 illustrates how our model can satisfactorily fit the measured Raman scattering data of Si NMs \cite{Lee2017}.

\begin{figure}[b]
\includegraphics[width=1.0\linewidth]{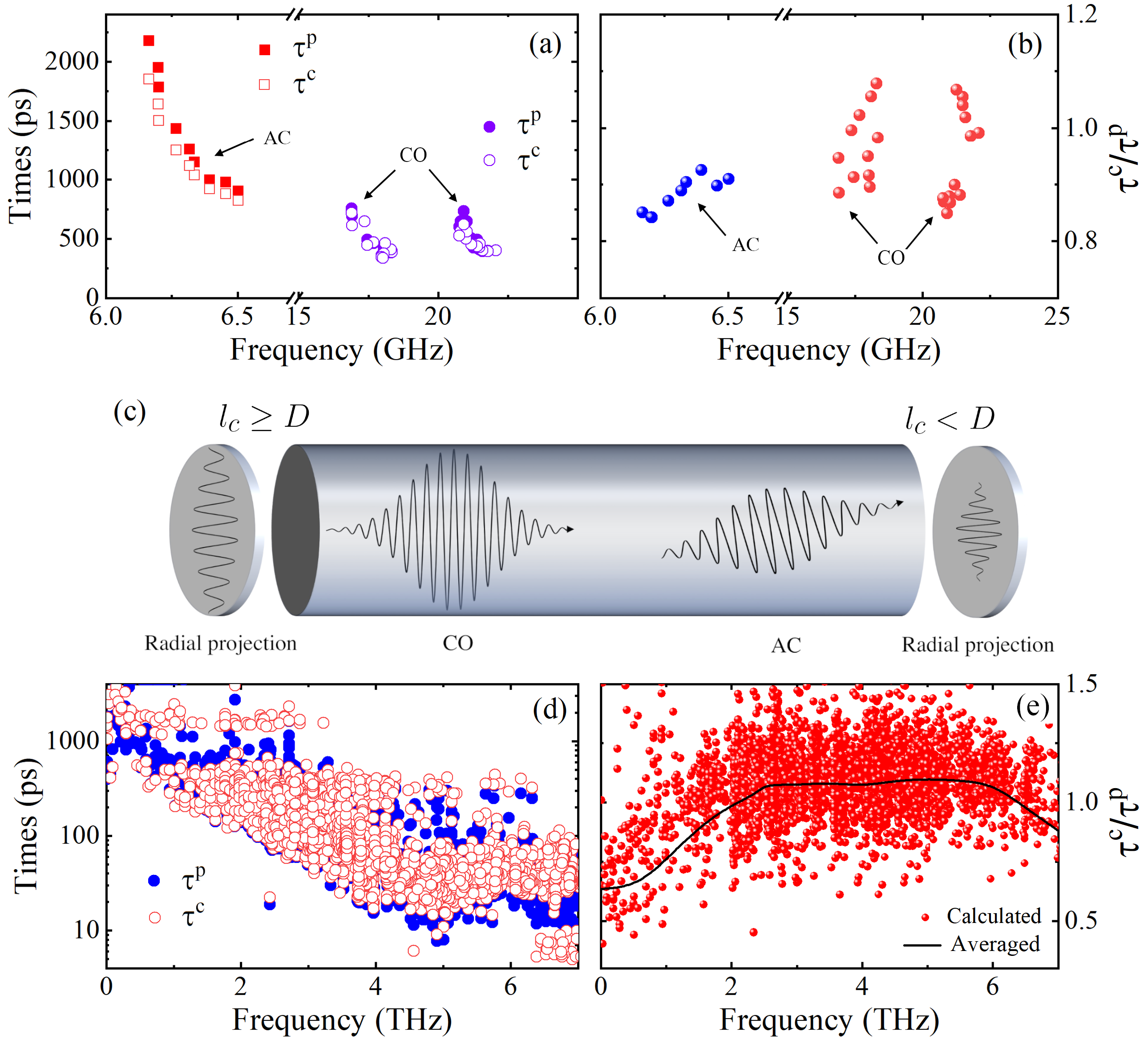}
\caption{(a) The estimated lifetimes ($\tau^{p}$) and coherence times ($\tau^{c}$) versus frequency obtained from the experimental spectroscopy of GaAs NWs with diameter $D=122$ nm. (b) The ratio $\tau^{c}/ \tau^{p}$ versus frequency for GaAs NWs. The AC and CO in (a) and (b) denote normal acoustic and confined optical modes, respectively. (c) A schematic of the different behaviors of AC and CO modes in NWs. $l_{c}$ is the size of the phonon wavepacket along the radial direction. $D$ refers to the nonperiodic (diameter) dimension of NWs. (d) The estimated lifetimes ($\tau^{p}$) and coherence times ($\tau^{c}$) versus frequency obtained from the MD spectral energy of Si NWs with diameter $D=1.5$ nm. (e) The ratio $\tau^{c}/ \tau^{p}$ versus frequency for Si NWs. The black line in (e) corresponds to the averaged ratio at a given frequency.}
\label{fig2}
\end{figure}

The estimated phonon lifetimes and coherence times of GaAs NWs are reported in Figs.\,\ref{fig2}(a) and \,\ref{fig2}(b), respectively. Lifetimes and coherence times are always close to each other and gradually decrease when frequency increases, which agrees qualitatively well with the general frequency dependence observed in various solids \cite{zhang2018reducing,Zhang2021,zhang2021heat,zhang2021strong}. The acoustic modes (AC) have higher characteristic times than those of confined optical modes (CO) at high frequencies.

We further evaluate the coherence of different modes through the ratio of coherence times to lifetimes, $\tau^{c}/ \tau^{p}$. This ratio compares the degree of the wavelike behavior to the particlelike one for a specific mode. Figure \,\ref{fig2}(b) highlights that when comparing the acoustic modes to the confined ones, the wavelike behavior of phonons gradually intensifies. As experimentally confirmed by Kargar $et$ $al.$ \cite{RN194}, those GaAs NWs modes are resulting from a confinement effect, which clearly exhibit a more pronounced coherence as compared to the bulk optical and the low-frequency AC modes. For CO modes, the size of the phonon wavepackets, i.e., the spatial coherence length of phonons $l_{c}$ \cite{RN1395,zhang2021strong}, along the radial direction is comparable or even larger than the non-periodic dimension, as indicated in Fig.\,\ref{fig2}(c). Consequently, phonon modes are confined along the diameter direction as standing waves with strong phonon coherence. Constratingly, the wavepackets of AC modes can tilt along any direction and the lengthes $l_{c}$ remain significantly smaller than the size of the non-periodic dimension $D$, yielding a predominant particlelike behaviour of those modes.

\begin{figure}[t]
\includegraphics[width=1.0\linewidth]{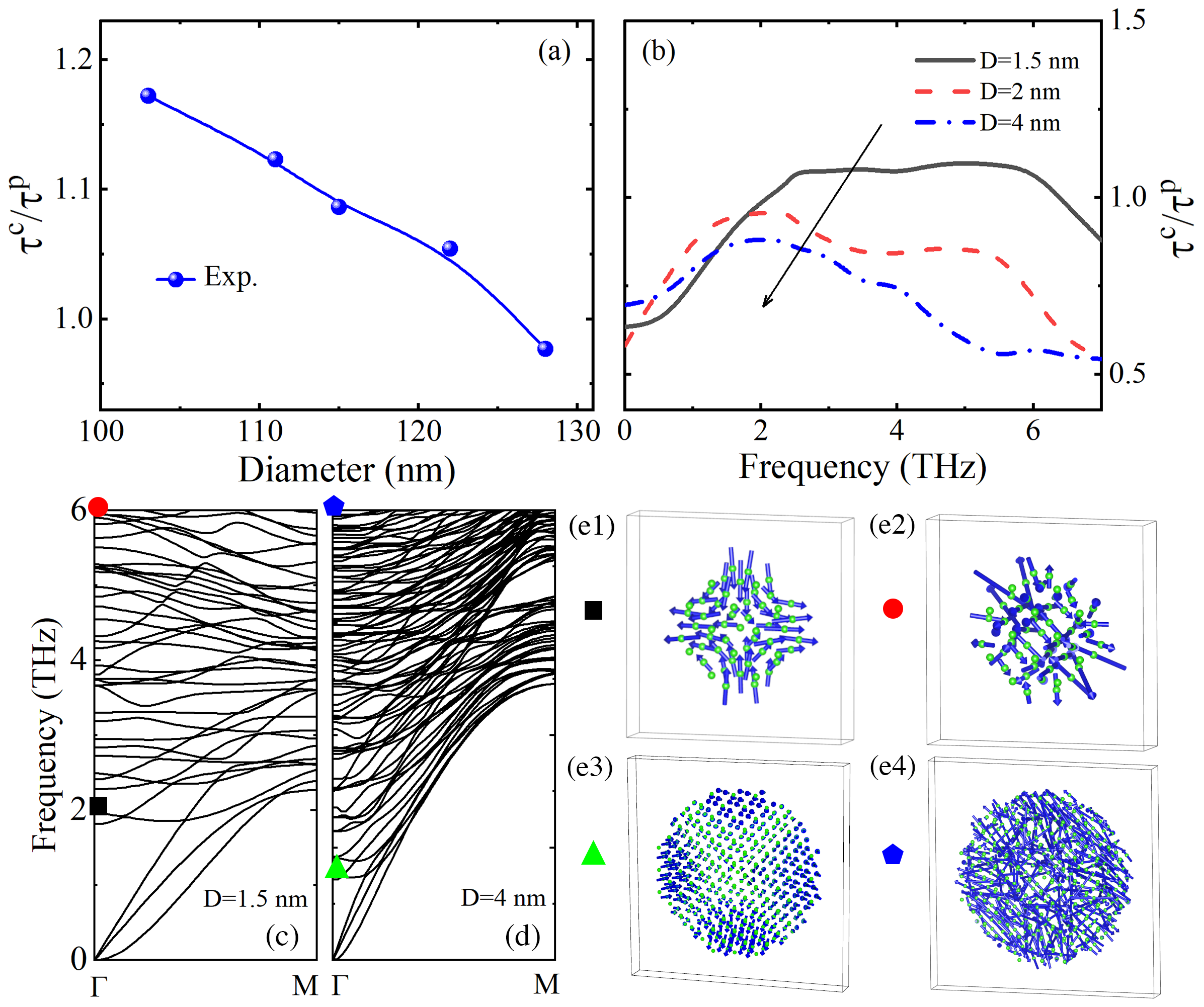}
\caption{(a) The ratio $\tau^{c}/ \tau^{p}$ versus diameter for the mode of lowest frequency in GaAS NWs. (b) The average ratio $\tau^{c}/ \tau^{p}$ versus frequency for Si NWs with different diameters. The arrow in (b) denotes an increasing trend with diameter. Phonon dispersion of Si NWs with diameters $D=1.5$ nm (c) and $D=4$ nm (d). (e1-4) Perspective views of the atomic eigen-displacements that correspond to the specific modes in (c) and(d) referred to as symbols. Arrows in (e1-4) denote the atomic vibrational directions.
}
\label{fig3}
\end{figure}

A further implementation of the proposed model to MD simulations is presented in Figs.\,\ref{fig2}(d) and \,\ref{fig2}(e). The decreasing trend of phonon lifetimes and coherence times as a function of frequency in the full Brillouin Zone is also obtained in Si NWs as a first validation. For phonons with frequencies below 2 THz, where the acoustic modes are predominant (See Fig.\,\ref{fig3}(c)), the values of $\tau^{c}/ \tau^{p}$ exhibit an increasing trend with frequency, indicating an enhanced coherence as approaching the confined modes frequency. This outcome qualitatively agrees with the results obtained from the spectroscopy of GaAs NWs.

Note that the full phonon dispersion in Fig.\,\ref{fig3}(c) contains a large range of modes which yields the scattered data of Fig.\,\ref{fig2}(e). The average (black line, averaged over frequency intervals) clearly discloses the enhanced wavelike behavior for confined modes in the middle frequency range. In addition, at frequencies above 6 THz, the ratio decreases with frequency, demonstrating that the high-frequency optical phonons resembles to bulk and weakly confined modes, as expected from modes with very small wavelengths.

The phonon confinement effect weakens when size in the confined direction increases, i.e. when diameter and thickness increase in NWs and NMs, respectively. The weakened confinement is also accompanied by a frequency reduction of the confined modes as observed in spectroscopy measurements \cite{RN194,Lee2017} and phonon dispersions of Figs.\,\ref{fig3}(c) and \,\ref{fig3}(d).

With the spectroscopy data of the GaAs NWs \cite{RN194}, we further estimate the dependence of the wavelike behavior on the diameter from the ratio $\tau^{c}/ \tau^{p}$. Figure \,\ref{fig3}(a) reports the decrease of the ratio $\tau^{c}/ \tau^{p}$ with increasing diameter for a confined mode, indicating the suppressed wavelike behaviors due to the weakened phonon confinement effect. The suppressed confinement with increasing thickness is also observed in two-dimensional Si NMs (See Sec. S3 in Supplementary Material \cite{SM}). 

We further study the size-dependent wavelike behavior using the MD spectral energy. Figure \,\ref{fig3}(b) highlights a pronounced impact of the diameter on the times ratio in Si NWs. The increase of $\tau^{c}/ \tau^{p}$ below 2 THz is shifted to a lower frequency as increasing the diameter due to the red-shift of the confined mode frequency (See Figs.\,\ref{fig3}(c) and \,\ref{fig3}(d)). Besides the suppression of the wavelike behavior around the frequency of the confined modes, a larger diameter substantially reduces the times ratio in the high frequency range, which might be caused by the stronger boundary scattering \cite{Martin2009}, yielding the simultaneous decrease of the two characteristic times.

To demonstrate phonon coherence, we further investigate the modal wave information of Si NWs from $Lattice$ $Dynamics$ calculations \cite{A606455H} in Figs. \,\ref{fig3}(c-e). Figs. \,\ref{fig3}(c-d) indicate that the confined modes around the region of lowest optical modes shift to lower frequencies as the diameter increases from 1.5 nm to 4 nm. In Fig. \,\ref{fig3}(e), we further compare the eigen-displacements of two kinds of optical modes at different frequencies (See the symbols in Figs. \,\ref{fig3}(c) and \ref{fig3}(d)), i.e., the confined optical modes in the low frequency region with high coherence, and the bulk optical modes with strong particlelike behavior. The eigen-displacements of the confined modes of the ultra-thin NWs (i.e., $D$ = 1.5 nm) reveal a strong confinement effect (See Fig. \,\ref{fig3}(e1)), in which the atoms at the four corners are collectively vibrating in phase and clearly interact with the boundary (See Video 1). However, for the $D=$ 4 nm NWs reported in Fig. \,\ref{fig3}(e3), the in-phase vibrations of the NW lattice cell
are still maintained but the interaction with boundary has been partially suppressed (See Video 2), resulting in a weakened confinement effect. The reduced wavelike behavior expressed by the eigen-displacements are well consistent with the estimated lifetimes and coherence times. 

On the other hand, the eigen-displacements of the bulk optical modes are also shown in Figs. \,\ref{fig3}(e2) and \,\ref{fig3}(e4). Random and disordered vibrations are found for these optical modes which are located in the high frequency region of the phonon dispersion. These vibrations with out-of-phase atomic oscillations indicate the weak wavelike behaviors as predicted by the proposed model in Figs.\,\ref{fig2}(e) and \,\ref{fig3}(b) (See Video 3 and 4).

Our formalism of mode energy delivers a description of phonon coherence and provides the measurable physical quantity of coherence time using conventional phonon spectroscopy. In the context of previous studies on nanostructures, we have demonstrated the general applicability of our theory in assessing phonon coherence from both experimental measurements and numerical simulations. As a general framework, the developed model is an effective approach for studying phonon coherence in diverse solids and provides a new metric to interpret broadly used spectroscopies.

\section*{\label{sec:level1}Acknowledgments}
 
We would like to thank Prof. Alexander A. Balandin and Prof. Fariborz Kargar of the University of California, Riverside for providing us the experimental spectroscopy data. This project is partially supported by the grants from the National Natural Science Foundation of China (Grant No. 12075168 and 11890703), and Science and Technology Commission of Shanghai Municipality (Grant No. 19ZR1478600 and 21JC1405600). This work is also supported in part by CREST JST (No. JPMJCR19I1 and JPMJCR19Q3).

\bibliographystyle{apsrev4-2}
\bibliography{library}

\end{document}

%\preprint{APS/123-QED}

\title{Supplementary Material for `Assessing Phonon Coherence Using Spectroscopy'}% Force line breaks with \\
  
\author{Zhongwei Zhang}
\affiliation{Center for Phononics and Thermal Energy Science, School of Physics Science and Engineering and China-EU Joint Lab for Nanophononics, Tongji University, Shanghai 200092, People's Republic of China}
\affiliation{Institute of Industrial Science, The University of Tokyo, Tokyo 153-8505, Japan}

\author{Yangyu Guo}
\affiliation{Institut Lumière Matière, Université Claude Bernard Lyon 1-CNRS, Université de Lyon, Villeurbanne 69622, France}

\author{Marc Bescond}
\affiliation{IM2NP, UMR CNRS 7334, Aix-Marseille Université, Faculté des Sciences de Saint Jérôme, Case 142, 13397 Marseille Cedex 20, France}

\author{Masahiro Nomura}
\affiliation{Institute of Industrial Science, The University of Tokyo, Tokyo 153-8505, Japan}

\author{Sebastian Volz}
\email{volz@iis.u-tokyo.ac.jp}
\affiliation{Center for Phononics and Thermal Energy Science, School of Physics Science and Engineering and China-EU Joint Lab for Nanophononics, Tongji University, Shanghai 200092, People's Republic of China}
\affiliation{Laboratory for Integrated Micro and Mechatronic Systems, CNRS-IIS UMI 2820, The University of Tokyo, Tokyo 153-8505, Japan}

\author{Jie Chen}
\email{jie@tongji.edu.cn}
\affiliation{Center for Phononics and Thermal Energy Science, School of Physics Science and Engineering and China-EU Joint Lab for Nanophononics, Tongji University, Shanghai 200092, People's Republic of China}

\maketitle

%\tableofcontents 

\section{Direct MD simulations}

\begin{figure}[h]
%[htb]
\includegraphics[width=0.6\linewidth]{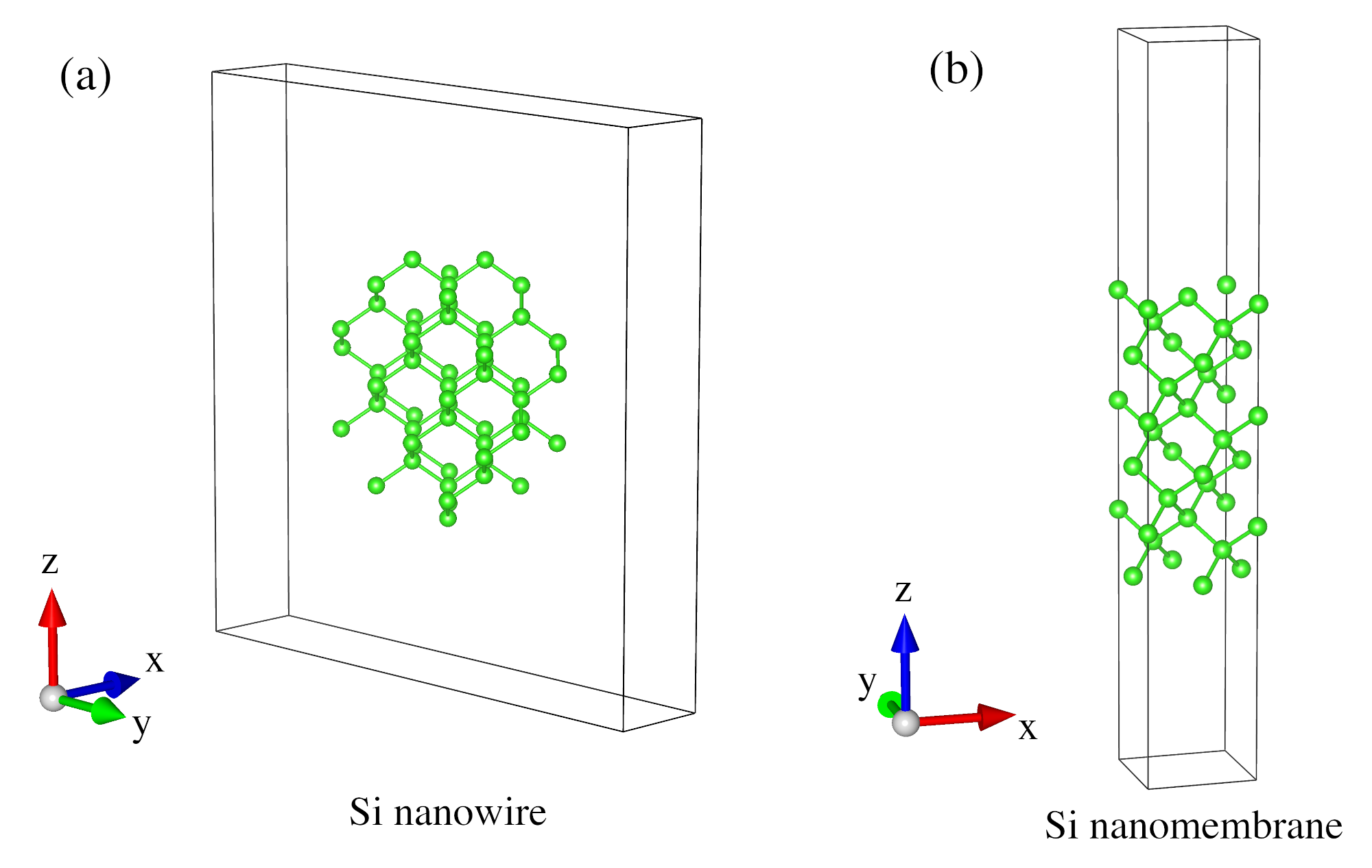}
\caption{(a) The atomic crystal of a Si nanowire with a 1.5 nm diameter. (b) The atomic crystal of a Si nanomembrane with a 1.5 nm thickness.
}
\label{figs1}
\end{figure}

All MD simulations are carried out using the LAMMPS package \cite{Plimpton1995} with a time step of 0.35 fs. The Si-Si interactions in silicon systems are modeled by the Stillinger-Weber potential \cite{Stillinger1985}. For the NWs, a sufficient vacuum space (15 $\AA$) is added to the $y$ and $z$ directions, and $x$ direction is simulated with a periodic boundary condition as shown by Fig.\,\ref{figs1}(a). For the NMs, $z$ direction contains 15 $\AA$ additional vacuum space, and $x$ and $y$ directions are simulated with a periodic boundary condition (See Fig.\,\ref{figs1}(b)). A 100$\times$1$\times$1 and 100$\times$20$\times$1 supercells are, respectively, applied to the MD simulations. After the structure relaxation and thermal equilibration in the isothermal-isobaric (NPT) ensemble for 500 ps, EMD simulations with the microcanonical (NVE) ensemble are performed to record the atomic trajectories.

\section{Normal mode decomposition}

The phonon modal velocity $\mathbf{\upsilon} \left ( \mathbf{k},s \right )$ can be expressed as \cite{Larkin2014a}

\begin{eqnarray}
\mathbf{\upsilon} \left ( \mathbf{k},s \right ) =\frac{1}{a}\sum_{b,l}^{a}\left [ \mathbf{\dot{u}}_{bl}\left ( t \right )\cdot \mathbf{e}^{\ast } _{b}\left ( \mathbf{k},s \right )\times exp\left ( i\mathbf{k}\cdot \mathbf{R}_{0l} \right )\right ],
\label{eqs1}
\end{eqnarray}

\noindent where $\mathbf{\dot{u}}_{bl}\left ( t \right )$ is the velocity of the $b$th atom in the $l$th unit cell at time $t$, $a$ is the number of cell, $\mathbf{e}^{\ast } \left ( \lambda \right )$ the complex conjugate of the eigenvector of mode $\lambda$, and $\mathbf{R}_{0l} $ is the equilibrium position of the $l$th unit cell. Here, $\mathbf{k}$ and $s$ correspond to the mode $\lambda$. The spectral energy can be further calculated as

\begin{eqnarray}
\Phi _{\mathbf{k}s}\left ( \omega  \right )=2\times \frac{1}{2} m\left | \int_{0}^{\infty } \mathbf{\upsilon} \left ( \mathbf{k},s \right ) e^{i\omega t}dt\right |^{2},
\label{eqs1}
\end{eqnarray}

\noindent where $m$ is the mass of the Si atom.

\clearpage

\section{Phonon coherence in Si nanomembrane}

\subsection{The fitting of spectroscopy data}

 \begin{figure}[h]
%[htb]
\includegraphics[width=0.5\linewidth]{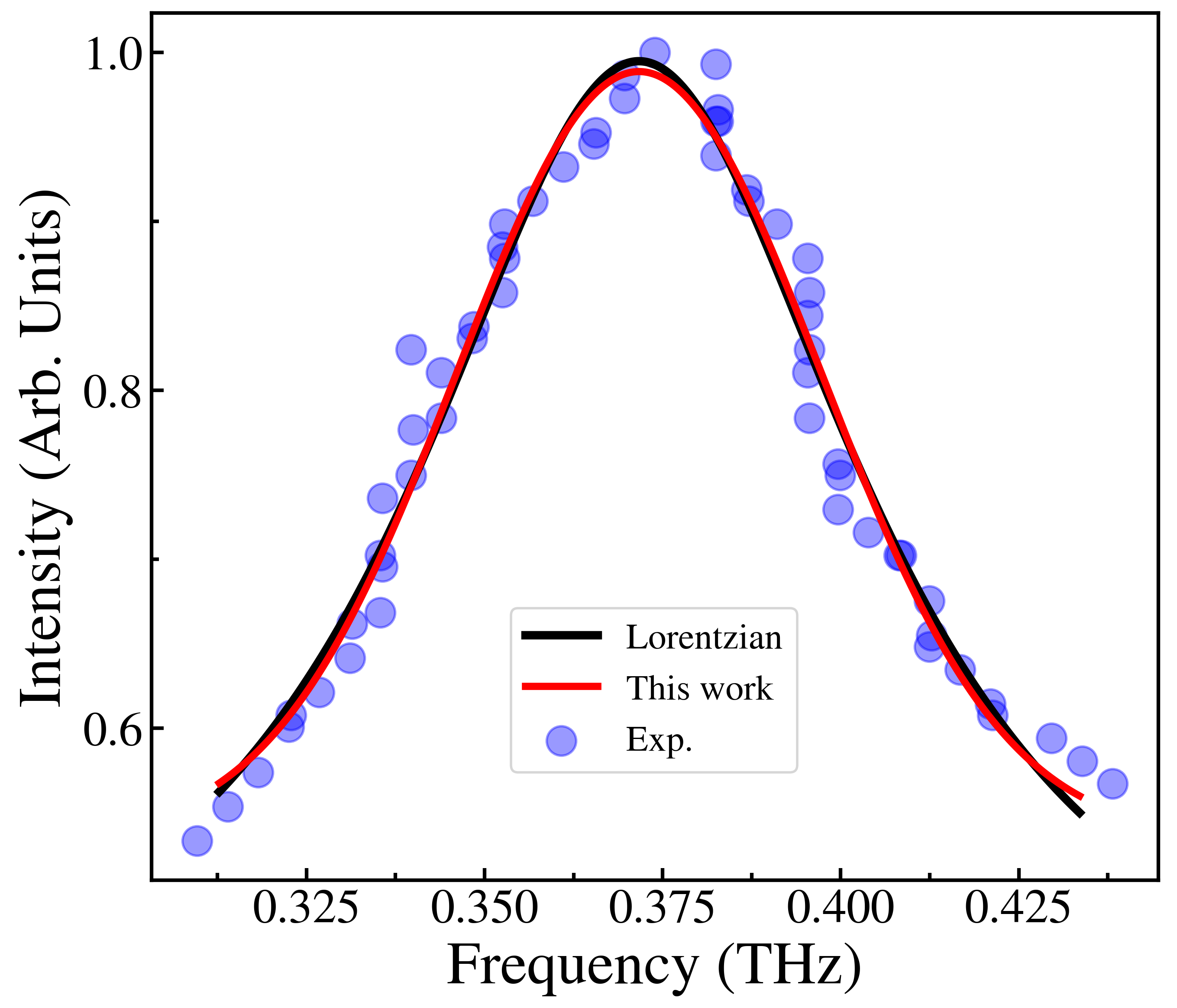}
\caption{The fitting of experimental spectroscopy with the classical Lorentzian model of Eq. (1) and with the proposed model of Eq. (6). The experimental spectroscopy data of a Si nanomembrane with thickness 8 nm were previously published by Lee $et$ $al.$ \cite{Lee2017}.
}
\label{figs2}
\end{figure}

~\\

\subsection{Thickness dependent phonon coherence}
     
 \begin{figure}[h]
%[htb]
\includegraphics[width=0.7\linewidth]{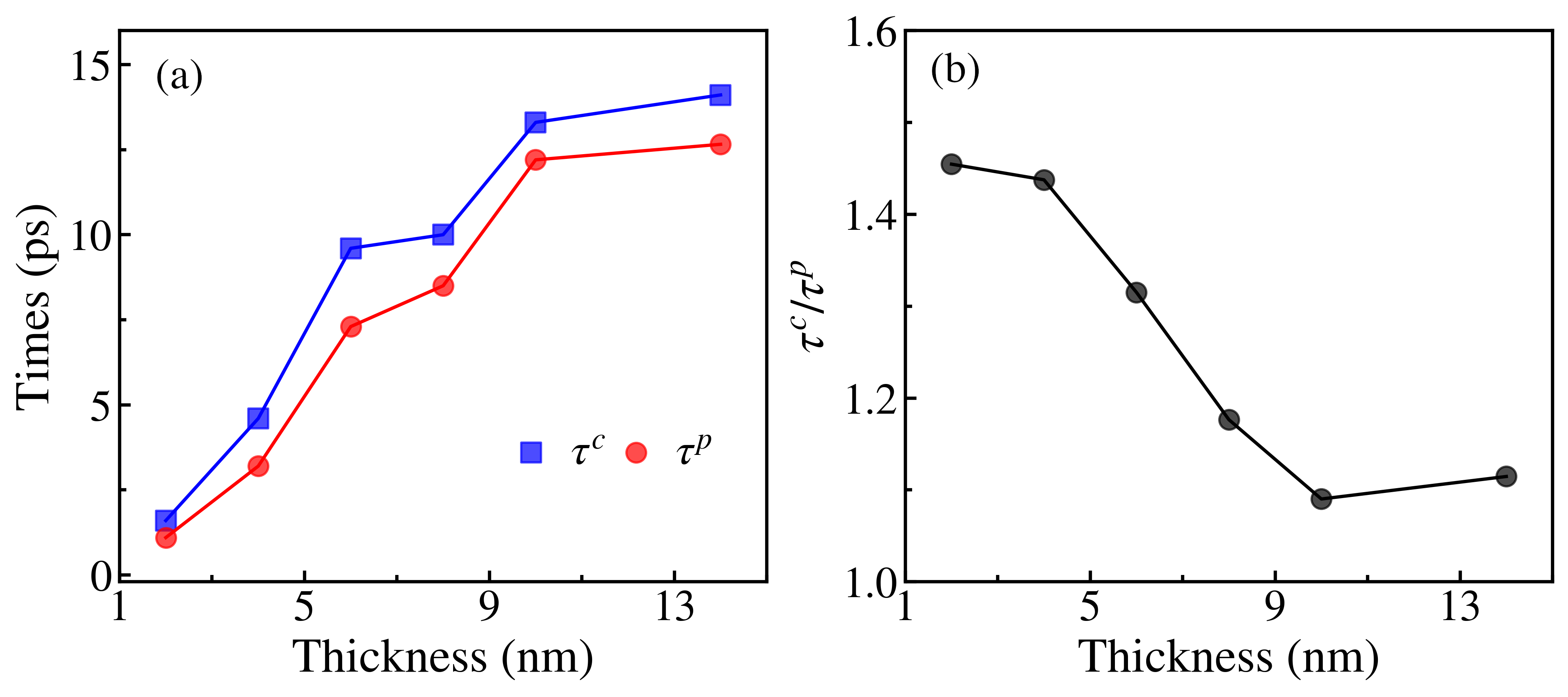}
\caption{(a) The estimated lifetimes ($\tau^{p}$) and coherence times ($\tau^{c}$) versus thickness obtained from the experimental spectroscopy data of Si nanomembranes. (b) The ratio $\tau^{c}/ \tau^{p}$ versus thickness for Si nanomembranes.
}
\label{figs3}
\end{figure}
 
~\\
~\\
~\\

\bibliographystyle{apsrev4-2}
\bibliography{library}